# Onset of Decoherence in Open Quantum Systems

Vladimir Privman

Center for Quantum Device Technology,
Department of Physics and Department of Electrical and Computer Engineering,
Clarkson University, Potsdam, New York 13699–5820, USA

Electronic mail: privman@clarkson.edu

## ABSTRACT

We introduce a new approximation scheme for evaluation of onset of decoherence at low temperatures in quantum systems interacting with environment. The approximation is argued to apply at short and intermediate times. It provides an approach complementary to Markovian approximations and appropriate for evaluation of quantum computing schemes.

## 1. INTRODUCTION

When a microscopic quantum system, $S$, with the Hamiltonian $H_S$, interacts with its environment, it is no longer described by a wavefunction. Rather, we have to use its density matrix, once the environment is traced over. Development of the ideas of quantum information, such as quantum computing (QC), spintronics, etc., has brought into focus and defined new questions in connection with concepts such as decoherence, thermalization, relaxation. Thus, our quantum system could be a single quantum two-state system (qubit) or it could be multi-qubit. Our presentation here is quite general.

In order to have controlled quantum dynamics, we aim at minimizing the effects of the environment. Therefore, we consider here those situations when the environmental effects are weak. Typically, interactions with the surrounding world are then quantified by identifying the modes of a bath, $B$, e.g., phonons, photons, spin-excitons, etc., which dominate the relaxation of the system $S$. Each bath mode is described by its Hamiltonian $M_K$, so that the bath Hamiltonian is

$$H_B = \sum_K M_K \ . \tag{1.1}$$

The interaction, $I$, of the bath modes with $S$, will be modeled by

$$H_I = \Lambda_S P_B = \Lambda_S \sum_K J_K ,  \qquad (1.2)$$

where $\Lambda_S$ is some Hermitean operator of $S$, coupled to the operator $P_B$ of the bath. A popular choice is the bath of bosonic (oscillator) modes,[1-6]

$$M_K = \omega_K a_K^\dagger a_K , \qquad (1.3)$$

$$J_K = g_K^* a_K + g_K a_K^\dagger . \qquad (1.4)$$

Here we use the units such that $\hbar = 1$. The total Hamiltonian of the system and bath is

$$H = H_S + H_B + H_I . \qquad (1.5)$$

More generally, the interaction, (1.2), can involve several system operators, each coupling differently to the bath modes, or even to different baths. The bath modes can be coupled to external objects, such as impurities, as well as interact with each other.

After the bath modes have been traced over, the system is described by the reduced density matrix, $\rho(t)$. If the system is not externally controlled, i.e., if $H_S$ is not time-dependent, then for large enough times we in principle expect thermalization. The density matrix should approach

$$\rho(t \to \infty) = \frac{\exp(-\beta H_S)}{\text{Tr}_S[\exp(-\beta H_S)]} , \qquad (1.6)$$

where $\beta \equiv 1/kT$. Actually, the model interactions (1.1)-(1.5) cannot yield thermalization without further Markovian assumption, which will be mentioned later. At times $t > 0$, the system deviates from coherent pure-quantum-state evolution. This departure is due to the interactions and entanglement with the bath. The temperature, $T$, and other external parameters that might be needed to characterize the system's density matrix, are determined by the properties of the bath, which in turn might interact with the rest of the universe.

We introduce the energy eigenstates, and the density-operator matrix elements,

$$H_S |n\rangle = E_n |n\rangle , \qquad (1.7)$$

$$\rho_{mn}(t) = \langle m | \rho(t) | n \rangle . \qquad (1.8)$$

As illustrated in Figure 1 (see next page), for large times we expect the diagonal elements $\rho_{nn}$ to approach values proportional to $e^{-\beta E_n}$, while the off-diagonal elements, $\rho_{m \neq n}$, to vanish. These properties can be referred to as thermalization and decoherence in the energy basis, and characterized by the time scales $T_1$ and $T_2$, respectively, though "thermalization," as defined by (1.6), implies decoherence.

In Section 2, we offer a survey of selected issues in studies of decoherence, Markovian approximation, and quantum computing. Then, in Section 3, we present our short-time-decoherence approximation. We offer arguments that, at low temperatures, this approximation is actually also valid for intermediate times and can be used to evaluate quantum computing designs. Results for the bosonic heat bath are given in Section 4. Section 5 addresses the case of adiabatic decoherence, when the short-time approximation becomes exact.

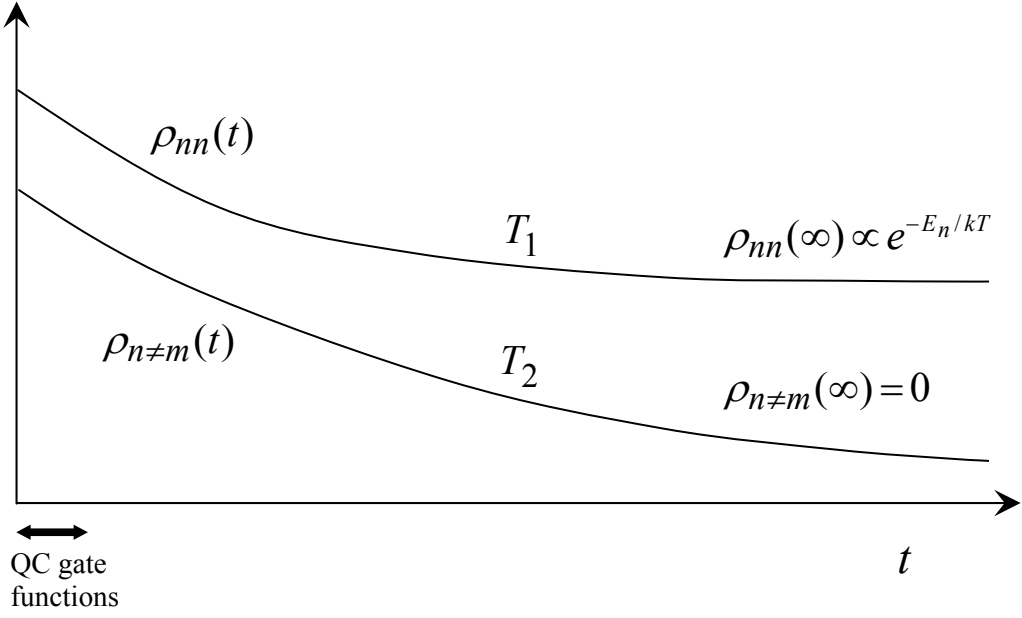

**Figure 1:** Behavior of the density matrix elements in the energy basis.

## 2. TIME SCALES OF RELAXATION AND QUANTUM COMPUTING

The behavior sketched in Figure 1, cannot be obtained within the model (1.1)-(1.5) without additional assumptions.[1-11] At time $t=0$, the bath modes, $K$, are assumed thermalized, i.e., have density matrices

$$\theta_K = e^{-\beta M_K} / \text{Tr}_K(e^{-\beta M_K}) \quad . \tag{2.1}$$

The density matrix $R$ of the system plus bath at time $t=0$ is assumed to be the direct product

$$R(0) = \rho(0) \otimes \theta_1 \otimes \theta_2 \otimes \cdots = \rho(0) \prod_K \theta_K \;, \tag{2.2}$$

where we will omit the direct product symbols, $\otimes$, from now on. A series of assumptions are made, e.g., the Markovian and secular approximations.[7-11] The most important is the Markovian approximation, which essentially assumes that the density matrices of the bath modes are reset externally to the thermal ones, on the time scale $\delta t$ shorter than any dynamical times of the system interacting with the bath, and the full density matrix is reset to the product form $\rho(t) \prod \theta_K$ after each time step $\delta t$. This is a natural assumption, because each bath mode is coupled only weakly to the system, whereas it is "monitored" by the rest of the universe and kept at temperature $T$. Ultimately, the aim is to derive master equations for the evolution of $\rho_{mn}(t)$, in the limit $\delta t \to 0$. The results are consistent with the Golden Rule and with the expected thermalization and decoherence properties; see Figure 1.

Several dynamical time scales can be identified. One is defined by the upper-frequency cutoff (Debye frequency) for the bath modes, $\omega_c$. This cutoff in the density of states need not be sharp, but it defines the time scale $1/\omega_c$. Usually, the time scales of the system's internal dynamics, $1/\Delta E_{mn}$, as well as the time scales that result from the system-bath interactions in the well-developed relaxation regime, $T_1$, $T_2$, etc., are larger than $1/\omega_c$. Note that these are the times that are experimentally observable, in an open quantum system, such as the "intrinsic" NMR/ESR relaxation times $T_1$ and $T_2$. Finally, there is the thermal time scale $\hbar/kT = \beta$; recall that we use units $\hbar = 1$.

The "resonant" bath modes with frequencies $\omega$ close to $\Delta E_{mn}$, can drive thermalization and the accompanying decoherence, by the system's actual emission and absorption of excitations to/from the bath. However, this typically yields $T_2$ values comparable to $2T_1$, whereas for most systems it is anticipated that $T_2 < T_1$. For systems of interest in quantum computing, we expect[12] that $T_2 \ll T_1$. Indeed, it can be argued that decoherence can result from processes that do not exchange energy between the system and bath, and thus it has more channels than resonant relaxation. This added "pure decoherence" results from virtual exchanges of excitations with the bath and is dominated by the bath modes near $\omega = 0$. Since it does not involve an internal energy scale, we can naturally associate[7,11,13] the thermal time $\beta$ with pure decoherence. It is the time it takes for the system's motion induced by its interactions with and at the frequencies of the low-energy bath modes, to "run out of phase," thus allowing the thermal fluctuation effects to exceed the purely quantum phase-decoherence effects.

There are indications[7,11,13] that the Markovian and similar approximations[7-11] used in many derivations of equations for thermalization and decoherence, are only valid for times larger than the thermal time scale $\beta$. We emphasize that not all the approximation schemes have this limitation;[6,11,14,15] one such approach is surveyed in this article.[14,15] We also point out that the above line of argument makes it tempting to identify the bath-recovery time scale $\delta t$, introduced earlier, with $\beta$. However, no definitive connection has been established thus far.[7-11]

For $T$ values close to room temperatures, ~ 300 K, we have $\hbar/kT \simeq 2.5 \cdot 10^{-14}$ sec. However, for quantum computing in solid-state semiconductor-heterostructure architectures,[12,16-22] we expect temperatures at least as low as several tens of mK. The thermal time scale is then between $10^{-10}$ and $10^{-9}$ sec, which is dangerously close to the external single-qubit control, Rabi-flip "quantum gate function" times even for the slowest qubits, those based on nuclear spins, which can be as short as $10^{-7}$ sec. Thus, for evaluation of quantum computing designs, new approximation schemes that do not utilize the conventional approximations are needed.

Quantum computing designs usually utilize systems, both the qubits and the modes that couple them, that have large spectral gaps. It is believed that, especially at low temperatures, spectral gaps slow down relaxation processes. Therefore, quantum computing architectures usually consider[16-22] qubits (two-state systems) in quantum dots, or in atoms, or subject to large magnetic fields, and coupled by highly nondissipative quantum media.[17,22] The spectral gaps are expected to slow down exponentially, by the Boltzmann factor, the processes of thermalization, involving energy exchange. Off-shell virtual exchanges, will be also slowed down, but less profoundly. The latter processes contribute to decoherence. Therefore, at low temperatures, we might expect separation of time scales of the initial decoherence vs. later-stage thermalization and further decoherence.

Since only the late-stage relaxation is clearly associated with the energy eigenbasis, we can pose the question whether the energy basis is the appropriate one to describe decoherence for short and intermediate times. In models of quantum measurement, it has been argued[23-27] that the eigenbasis of the interaction operator, $\Lambda_S$, may be more appropriate. Thus, in addition to the energy basis, (1.7), we also define the eigenstates of the interaction operator $\Lambda_S$, by

$$\Lambda_S |\gamma\rangle = \lambda_\gamma |\gamma\rangle , \qquad (2.3)$$

where from now on the Greek letters will label the eigenstates of $\Lambda_S$, with eigenvalues $\lambda_\gamma$, while the Roman letters will be used for the energy eigenbasis, (1.7), and, when capitalized, for the bath modes, (1.2)-(1.4). Ideally, we would like to have basis-independent approximations for the operator $\rho(t)$.

The quantum computation process[28-33] necessitates a succession of gate functions, whereby $H_S$ of a multiqubit system is "controlled," i.e., becomes time-dependent, and also error correction that might involve measurement of some of the qubits. Therefore, our model (1.1)-(1.5), with constant $H_S$ and presumably only few qubits to make the calculations tractable, can at best be used to evaluate the degree of decoherence for times comparable to single gate functions; see Figure 1. The quantum error correction criterion can be then tested: the error rate should be somewhere between $10^{-6}$ to $10^{-4}$, depending on the system under consideration.[28-33] Variation of the matrix elements of $\rho(t)$, in whatever basis, might not be the best measure of the degree of relaxation. Other measures, derivable from the density matrix, have been proposed.[34] In this article, we will use the degree of departure from a pure state, measured by the deviation of $\text{Tr}_S[\rho^2(t)]$ from 1.

Recently, there have been several calculations of spin decoherence in solid state systems appropriate for quantum computing.[13,22,27,35-50] Some of these works have not invoked the traditional approximations, or have relied on the spectral gap of the bath modes, and included interactions of the latter with impurities,[40,45] to achieve better reliability of the results at low temperatures. In the present work, we consider bath modes interacting only with the system.

In order to better understand relaxation processes in situations when energy exchange with the bath is negligible, we have proposed an approach termed adiabatic decoherence,[27] extending the earlier works.[13,35-37,51] Thus, we assume that $H_S$ is conserved (a variant of a quantum nondemolition process),

$$[H_S, H] = [H_S, \Lambda_S] = 0 \quad \text{(adiabatic assumption).} \qquad (2.4)$$

This assumption precludes energy exchange, artificially leaving only energy-conserving relaxation pathways that contribute to decoherence. We will comment on the results of this approach in Section 5. Certain models of quantum measurement[52,53] evaluate decoherence by effectively setting $H_S = 0$, which is a special case of (2.4). Our approximation scheme described in Sections 3-4, is exact for the adiabatic case.

The formulation in Section 3, will be quite general. However, we do utilize the factorization property (2.2) at time $t = 0$. Thus, we do have to assume that, at least initially, the system and bath modes are not entangled. This factorization assumption, shared by all the recent spin-decoherence studies, represents the expectation that external quantum-gate-function control by short-duration but large externally applied potentials, measurement, etc., resets the qubits, disentangling them from the environmental modes to which the affected qubits are only weakly coupled. Thus, in quantum computing, it is the qubit system that gets approximately reset and disentangled from the bath towards time $t = 0$, instead of the bath being initially thermalized by the rest of the universe, as assumed in traditional formulations of quantum relaxation.

## 3. INITIAL DECOHERENCE

The time dependence of the overall density matrix $R(t)$ of the system and bath, is given by

$$R(t) = e^{-i(H_S + H_B + H_I)t} R(0) e^{i(H_S + H_B + H_I)t} \quad . \tag{3.1}$$

As our short-time approximation, we utilize the following approximate relation, expressing the exponential factors is (3.1) as products of unitary operators,

$$e^{i(H_S + H_B + H_I)t + O(t^3)} = e^{iH_S t/2} e^{i(H_B + H_I)t} e^{iH_S t/2} \quad . \tag{3.2}$$

Relations of this sort have been widely used in Field Theory[54] and Statistical Mechanics.[55] Specifically, (3.2) has the following appealing properties. It becomes exact for the adiabatic case, (2.4). Furthermore, if we use the right-hand side and its inverse to replace $e^{\pm iHt}$, then we are imposing three time-evolution-type transformations on $R(0)$. Therefore, the approximate expression for $R(t)$ will have all the desired properties of a density operator. Finally, extensions to higher-order approximations in powers of $t$ are possible, though cumbersome; expressions valid to $O(t^4)$ and $O(t^5)$ have been tabulated.[55,56]

We now consider the approximation to the matrix element,

$$\rho_{mn}(t) = \text{Tr}_B \langle m | e^{-iH_S t/2} e^{-i(H_B + H_I)t} e^{-iH_S t/2} R(0) e^{iH_S t/2} e^{i(H_B + H_I)t} e^{iH_S t/2} | n \rangle \quad . \tag{3.3}$$

We would like to replace all the system operators in (3.3) by numbers, at the expense of introducing summations. We apply the outer $H_S$'s to the left on $\langle m |$, and to the right on $| n \rangle$, replacing $H_S$ by, respectively, $E_m$ and $E_n$. Since the $H_I$'s in the next two exponential operators contain $\Lambda_S$, see (1.2), we insert the decomposition of the unit operator in the system space, in terms of the eigenbasis of $\Lambda_S$, before the second exponential, and after the penultimate one. We also insert the decomposition of the unit operator in the eigenbasis of $H_S$, before/after the two inner exponentials, which contain $H_S$. We get the result

$$\rho_{mn}(t) = \sum_{\gamma p q \delta} \text{Tr}_B [e^{-iE_m t/2} \langle m | \gamma \rangle \langle \gamma | p \rangle e^{-i(H_B + \lambda_\gamma P_B)t} e^{-iE_p t/2} \rho_{pq}(0)$$

$$\times (\prod_K \theta_K) e^{iE_q t/2} e^{i(H_B + \lambda_\delta P_B)t} \langle q | \delta \rangle \langle \delta | n \rangle e^{iE_n t/2} ] .$$

(3.4)

The key observation is that, with (1.1)-(1.2), the terms in (3.4) can be rearranged in such a way that the trace over the bath can be carried out for each mode separately,

$$\rho_{mn}(t) = \sum_{\gamma p q \delta} \{ e^{i(E_q + E_n - E_p - E_m)t/2} \langle m | \gamma \rangle \langle \gamma | p \rangle \rho_{pq}(0) \langle q | \delta \rangle \langle \delta | n \rangle$$

$$\times \prod_K \text{Tr}_K [e^{-i(M_K + \lambda_\gamma J_K)t} \theta_K e^{i(M_K + \lambda_\delta J_K)t}] \} .$$

(3.5)

  For the simplest quantum-computing application involving a single qubit, the four sums in (3.5) are over two terms each. The calculations of the overlap Dirac brackets between the eigenstates of $H_S$ (labeled by $m$, $n$, $p$ and $q$) and those of $\Lambda_S$ (labeled by $\gamma$ and $\delta$), as well as the energy-basis matrix elements of $\rho(0)$, involve at most diagonalization of two-by-two Hermitean matrices. Of course, the approximation (3.5) can be used for evaluation of short-time density matrices for systems more general than two-state. A challenging part of the calculation is the trace over each mode of the bath. Since these modes have identical structure, e.g., (1.3)-(1.4) for the bosonic bath case, but with $K$-dependent coupling constants, the calculation needs only be done once, in the space of *one mode*.

  An important question in connection with the approximation (3.5) is why don't we expand directly in powers of the time, $t$? Why use the approximation (3.2)? To address this issue, let us consider dimensionless combinations which can be constructed from $t$ and the characteristic frequencies of the problem, mentioned in Section 2. A brute-force expansion in powers of $t$ would involve the combination $\omega_c t$, and is expected to hold up to times $1/\omega_c$. Consideration of the form of corrections[55,56] to (3.2), reveals that they involve various commutators constructed from the operators in the exponents on the right-hand side of (3.2), namely, $H_S$ and $H_B + H_I$. Here only $H_B$ depends on the bath-mode frequencies and can introduce the cutoff frequency dependence. However, $H_B$ will drop out of any commutators, because it commutes with $H_S$. While not rigorous, this argument suggests that the higher-order corrections involve dimensionless combinations of $t$ with quantities constructed out of the energies of the system, e.g., the gaps $\Delta E_{mn}$, and those entering in the interaction operator, $\lambda_\gamma | g_K |$, see (1.2) and (1.4). Thus, we expect that for low temperatures, when the time scale of the fully developed thermalization, $\beta$, is the largest in the problem, our approximation is valid beyond the short times, defined by $t < O(1/\omega_c)$.

The approximation should hold up to intermediate times defined by the energy differences of the system Hamiltonian and interaction operator. It definitely breaks down for times of order $\beta$.

## 4. THE BOSONIC HEAT BATH

In this section, we consider the bosonic heat bath,[6] see (1.3)-(1.4), in the initially thermalized state, for which (2.1) gives

$$\theta_K = \left(1 - e^{-\beta \omega_K}\right) e^{-\beta \omega_K a_K^\dagger a_K} . \tag{4.1}$$

The product of the single-mode traces in (3.5), is actually available in the literature,[13,27,35]

$$\rho_{mn}(t) = \sum_{\gamma p q \delta} \{ e^{i(E_q + E_n - E_p - E_m)t/2} \langle m | \gamma \rangle \langle \gamma | p \rangle \langle q | \delta \rangle \langle \delta | n \rangle \rho_{pq}(0)$$

$$\times \exp\left(-\sum_K \frac{|g_K|^2}{\omega_K^2}\left[2(\lambda_\gamma - \lambda_\delta)^2 \sin^2\frac{\omega_K t}{2}\coth\frac{\beta \omega_K}{2} + i(\lambda_\gamma^2 - \lambda_\delta^2)(\sin \omega_K t - \omega_K t)\right]\right)\} . \tag{4.2}$$

The last term in the exponent, linear in $t$, can be viewed as "renormalization" of the system's energy due to the interaction with the bath. It can be removed by adding the term $H_R = \Lambda_S^2 \sum |g_K|^2 / \omega_K$ to the total Hamiltonian. However, the usefulness of this identification for short times is not clear, and we will not employ it.

We define two non-negative real spectral sums over the bath modes,

$$B^2(t) = 8 \sum_K \frac{|g_K|^2}{\omega_K^2} \sin^2 \frac{\omega_K t}{2} \coth \frac{\beta \omega_K}{2} , \tag{4.3}$$

$$C(t) = \sum_K \frac{|g_K|^2}{\omega_K^2}(\omega_K t - \sin \omega_K t) . \tag{4.4}$$

In the continuum limit of infinitely many bath modes, these sums have been discussed extensively in the literature,[6,13,35] for several choices of the bath-mode density of states and coupling strength, $g(\omega)$, as functions of the mode frequency. Relation (4.2) can now be written

$$\rho_{mn}(t) = \sum_{\gamma p q \delta} \{ e^{i(E_q + E_n - E_p - E_m)t/2} \langle m | \gamma \rangle \langle \gamma | p \rangle \langle q | \delta \rangle \langle \delta | n \rangle \rho_{pq}(0)$$

$$\times \exp[-\frac{1}{4} B^2(t)(\lambda_\gamma - \lambda_\delta)^2 + i C(t)(\lambda_\gamma^2 - \lambda_\delta^2)] \} . \tag{4.5}$$

We can derive a basis-independent representation for $\rho(t)$ by utilizing the identity

$$\sqrt{\pi}\exp[-B^2(\Delta\lambda)^2/4] = \int_{-\infty}^{\infty} dy\, e^{-y^2} \exp[iyB(\Delta\lambda)] \ . \tag{4.6}$$

Exponential factors in (4.5) can now be converted back to operators acting on the wavefunctions entering the overlap Dirac brackets, with the result

$$\sqrt{\pi}\,\rho(t) = \int_{-\infty}^{\infty} dy\, e^{-y^2} e^{-iH_S t/2} e^{i[yB(t)\Lambda_S + C(t)\Lambda_S^2]} e^{-iH_S t/2} \rho(0)\, e^{iH_S t/2} e^{-i[yB(t)\Lambda_S + C(t)\Lambda_S^2]} e^{iH_S t/2}. \tag{4.7}$$

This basis-independent expression also makes the deviation from a pure state $\rho(0) = |\psi_0\rangle\langle\psi_0|$ apparent: $\rho(t>0)$ is then obviously a *mixture* (integral over $y$, with $e^{-y^2}/\pi^{1/2}$ weight factor) of pure-state projectors $|\psi(y,t)\rangle\langle\psi(y,t)|$, where

$$|\psi(y,t)\rangle = e^{-iH_S t/2} e^{i[yB(t)\Lambda_S + C(t)\Lambda_S^2]} e^{-iH_S t/2} |\psi_0\rangle \ . \tag{4.8}$$

As an application, let us consider the case of $H_S$ proportional to the Pauli matrix $\sigma_z$, e.g., a spin-1/2 particle in magnetic field, and $\Lambda_S = \sigma_x$. We measure the deviation of the state of this qubit, assumed initially in the energy eigenstate $|\uparrow\rangle$ or $|\downarrow\rangle$, from a pure state, by calculating $\text{Tr}_S[\rho^2(t)]$ according to (4.7), with the result

$$\text{Tr}_S[\rho^2(t)] = \frac{1}{2}[1 + e^{-2B^2(t)}] \ . \tag{4.9}$$

For a two-state density matrix, $\text{Tr}(\rho^2)$ can vary from 1 for pure quantum states to the lowest value of 1/2 for maximally mixed states. Generally, as the time increases, the function $B^2(t)$ grows monotonically from zero.[6,13,27,35] Specifically, for Ohmic dissipation, $B^2(t)$ is known to increase quadratically for short times $t < O(1/\omega_c)$, then logarithmically for $O(1/\omega_D) < t < O(\hbar/kT)$, and linearly for $t > O(\hbar/kT)$. Our approximation yields reasonable results in the first two regimes. However, it cannot give thermalization in the regime $t > O(\beta)$. Instead, it predicts approach to the maximally mixed state. For other bath models, $B^2(t)$ need not diverge to infinity at large times.[6,13,27,35]

## 5. COMMENTS ON THE ADIABATIC CASE AND SUMMARY

The adiabatic assumption (2.4) corresponds to the system's energy conservation. Therefore, energy flow in and out of the system is not possible, and normal thermalization mechanisms are blocked. This "adiabatic decoherence" limit thus corresponds to pure dephasing.[27,51] Our approximation becomes exact in this case. Indeed, we can select a common eigenbasis for $H_S$ and $\Lambda_S$. The overlap Dirac brackets in (3.5) then become Kronecker symbols, and the sums can be evaluated to yield

$$\rho_{mn}(t) = e^{i(E_n - E_m)t} \rho_{mn}(0) \prod_K \text{Tr}_K [e^{-i(M_K + \lambda_m J_K)t} \theta_K e^{i(M_K + \lambda_n J_K)t}] \ . \tag{5.1}$$

This expression was discussed in detail in our work on adiabatic decoherence.[27] Specifically, for the initially thermalized bosonic heat bath case, we have, for the absolute values of the density matrix elements,

$$|\rho_{mn}(t)| = |\rho_{mn}(0)| e^{-B^2(t)(\lambda_m - \lambda_n)^2/4} \ . \tag{5.2}$$

The decay of the off-diagonal matrix elements is determined by the spectral function $B^2(t)$.

In summary, we have derived a new approximation for the density matrix. The expressions are easy to work with, because for few-qubit systems they only involve manipulation of finite-dimensional matrices, and they will be useful in estimating decoherence and deviation from pure states in quantum computing systems, specifically results for short and intermediate times, at low temperatures.

This research was supported by the National Science Foundation, grants DMR-0121146 and ECS-0102500, and by the National Security Agency and Advanced Research and Development Activity under Army Research Office contract DAAD-19-02-1-0035.